\documentclass[conference,9pt]{IEEEtran}
\IEEEoverridecommandlockouts
\usepackage{cite}
\usepackage{amsmath,amssymb,amsfonts}
\usepackage{graphicx}
\usepackage{textcomp}
\usepackage{amsmath}
\usepackage{booktabs} 
\usepackage{multirow}
\usepackage[font=footnotesize,skip=0pt]{caption}
\usepackage{setspace}
\usepackage{tabularx}
\usepackage{hyperref}       % hyperlinks
\usepackage{url}            % simple URL typesetting
\usepackage{booktabs}       % professional-quality tables
\usepackage{amsfonts}       % blackboard math symbols
\usepackage{nicefrac}       % compact symbols for 1/2, etc.
\usepackage{microtype}      % microtypography
\usepackage{xcolor} 
\usepackage{colortbl}
\usepackage{svg}
\usepackage{adjustbox}

\usepackage{amsmath}  % For mathematical expressions
\usepackage[linesnumbered,ruled,vlined]{algorithm2e}

\usepackage{booktabs}
\usepackage{amsmath}
\usepackage{tikz}

\usepackage{orcidlink}
\usepackage{comment}
\usepackage{amssymb}
\usepackage{braket}
\usepackage{xcolor}
\usepackage{cleveref}
\def\BibTeX{{\rm B\kern-.05em{\sc i\kern-.025em b}\kern-.08em
    T\kern-.1667em\lower.7ex\hbox{E}\kern-.125emX}}
\begin{document}

\title{QUAV: Quantum-Assisted Path Planning and Optimization for UAV Navigation with Obstacle Avoidance}

%\author{\IEEEauthorblockN{Anonymous Authors}}
%\begin{comment}
\author{\IEEEauthorblockN{Nouhaila Innan\textsuperscript{1,2}, Muhammad Kashif\textsuperscript{1,2}, Alberto Marchisio\textsuperscript{1,2}, Yung-Sze Gan\textsuperscript{3}, Frederic Barbaresco\textsuperscript{4}, Muhammad Shafique\textsuperscript{1,2}
\IEEEauthorblockA{
\textsuperscript{1}eBRAIN Lab, Division of Engineering, New York University Abu Dhabi (NYUAD), Abu Dhabi, UAE\\
\textsuperscript{2}Center for Quantum and Topological Systems (CQTS), NYUAD Research Institute, NYUAD, Abu Dhabi, UAE\\
\textsuperscript{3}Thales Solutions Asia Pte. Ltd., Singapore \\
\textsuperscript{4}Thales Land \& Air Systems, Thales Group, Paris, France \\
nouhaila.innan@nyu.edu, muhammadkashif@nyu.edu, alberto.marchisio@nyu.edu, \\yungsze.gan@asia.thalesgroup.com, frederic.barbaresco@thalesgroup.com, muhammad.shafique@nyu.edu,\\
}}}
%\end{comment}

\maketitle

\begin{abstract}
The growing demand for drone navigation in urban and restricted airspaces requires real-time path planning that is both safe and scalable. Classical methods often struggle with the computational load of high-dimensional optimization under dynamic constraints like obstacle avoidance and no-fly zones. This work introduces QUAV, a quantum-assisted UAV path planning framework based on the Quantum Approximate Optimization Algorithm (QAOA), to the best of our knowledge, this is one of the first applications of QAOA for drone trajectory optimization. QUAV models pathfinding as a quantum optimization problem, allowing efficient exploration of multiple paths while incorporating obstacle constraints and geospatial accuracy through UTM coordinate transformation.
A theoretical analysis shows that QUAV achieves linear scaling in circuit depth relative to the number of edges, under fixed optimization settings. Extensive simulations and a real-hardware implementation on IBM's \texttt{ibm\_kyiv} backend validate its performance and robustness under noise. Despite hardware constraints, results demonstrate that QUAV generates feasible, efficient trajectories, highlighting the promise of quantum approaches for future drone navigation systems.
\end{abstract}

%\begin{spacing}{0.975}
\section{Introduction}
%incomplete

The increasing complexity of urban airspaces and the rapid growth of Unmanned Aerial Vehicle (UAV) operations demand innovative solutions for real-time trajectory optimization. Traditional path planning methods~\cite{debnath2019review} struggle with the computational burden of evaluating vast numbers of potential routes in environments laden with obstacles and dynamic constraints. This paper addresses the research problem of efficiently computing collision-free, optimal trajectories for drones operating in complex, data-rich environments.

One major challenge is managing the computational complexity inherent in high-dimensional optimization problems. As the number and diversity of obstacles increase, the search space expands dramatically, often rendering classical algorithms impractical for large-scale scenarios. Moreover, the integration of heterogeneous geospatial data, ranging from varying coordinate systems to diverse environmental constraints, compounds the challenge. The process of converting and assimilating such data into a coherent model that accurately represents the operational environment requires meticulous preprocessing and transformation.

To overcome these challenges, we propose QUAV, a novel quantum-assisted framework that reformulates UAV path planning as a quantum optimization task using the Quantum Approximate Optimization Algorithm (QAOA) \cite{blekos2024review}. Our approach integrates theoretical analysis, simulation-based evaluation, and real-hardware implementation. The key contributions of this study are as follows:

\begin{itemize}
    \item \textbf{Formulation of a QAOA-based path planning framework:} We develop QUAV, a quantum-assisted trajectory optimization framework that integrates obstacle avoidance constraints into the quantum optimization process. This represents the first application of QAOA in the context of drone path planning.

    \item \textbf{Theoretical complexity analysis:} We analyze the computational complexity of QUAV, demonstrating its linear scaling with the number of edges. This provides insights into its computational feasibility compared to conventional path planning methods.

    \item \textbf{Simulation-based performance evaluation:} We evaluate QUAV’s behavior across multiple path planning scenarios, analyzing its ability to generate feasible and efficient paths under varying environmental constraints.

    \item \textbf{Implementation and execution on quantum hardware:} We implement QUAV on IBM’s \texttt{ibm\_kyiv} QPU to assess the impact of real quantum hardware constraints, including noise effects and qubit connectivity limitations, on the optimization process.
\end{itemize}

The rest of the paper is organized as follows: Sec.~\ref{sec2} provides an overview of the related work, discussing classical and quantum approaches to path planning. Sec.~\ref{sec22} introduces the problem formulation. Sec.~\ref{sec3} details the proposed methodology, including the QAOA-based formulation and implementation of QUAV. Sec.~\ref{sec4} presents the results, analyzing the performance in terms of path efficiency, computational complexity, and hardware execution. Sec.~\ref{sec5} provides a discussion, interpreting key findings and explaining observed patterns in the optimization behavior. Finally, Sec.~\ref{sec6} summarizes our findings and highlights potential directions for future research.

\section{Background and Related Work \label{sec2}}
The problem of autonomous path planning for UAVs has garnered significant attention across various domains, including dense urban transportation, emergency response, and military surveillance. The overarching goal is to compute safe and efficient paths that minimize key performance metrics such as flight time, energy consumption, or distance traveled, all while respecting obstacles and dynamic threats. Classic strategies for path planning often revolve around sampling-based approaches (e.g., Rapidly-exploring Random Tree) \cite{elbanhawi2014sampling,zhang2025motion}, graph-search methods (e.g., A$^\ast$, Dijkstra) \cite{debnath2019review}, or nature-inspired optimization (e.g., swarm intelligence) \cite{ayawli2018overview,jiang2024evolutionary}. More recently, deep reinforcement learning has been adopted to tackle high-dimensional search spaces and environmental uncertainties \cite{zhang2024recent}. Despite considerable progress, large-scale or real-time scenarios continue to pose computational and algorithmic challenges.  

\subsection{Classical Approaches}
Among deterministic and learning-based approaches, wind disturbances, obstacle avoidance, and energy efficiency are major design factors \cite{reda2024path}. An improved Double Deep Q-Network (DDQN) incorporating wind in dense urban environments has been proposed to enhance both autonomy and adaptability of UAVs \cite{zhu2024uav}. Meanwhile, swarm optimization algorithms, such as Ant Colony Optimization (ACO) and Particle Swarm Optimization (PSO), have been integrated into multi-drone path planners, effectively balancing collision avoidance and travel time by means of social-agent heuristics \cite{saeed2022optimal}. 

Other investigations highlight real-time collision-free motion in uncertain or partially known environments. For instance, robust path-planning techniques based on mathematically proven collision avoidance have been shown to reduce on-board computational loads by focusing only on interest points near rectangular obstacles \cite{bashir2023obstacle}. Furthermore, 3D voxel-based methods paired with jump point search (JPS) can systematically account for static obstacles, whereas local collision-resolution schemes address newly detected threats in real time \cite{luo20223d}. Similar progress has been reported on multi-drone coordination, wherein offline algorithms such as DETACH and STEER iteratively minimize intersections of flight paths to avert collisions \cite{shen2022multidepot}. Evolutionary strategies, such as improved sparrow search algorithms \cite{he2025dynamic}, have also been integrated to mitigate randomness and accelerate convergence in UAV path planning, with extended applications to ant-colony-based frameworks for adaptive parameter updates in dynamic environments \cite{zhang2025uav}.
\subsection{Quantum-Assisted Path Planning}
Motivated by the combinatorial nature of path-planning problems, recent research has begun exploring quantum and hybrid quantum-classical approaches. For instance, variants of Bloch Spherical Quantum Genetic Algorithms (BQGA) and Quantum Bee Colony Algorithms (QABC) have demonstrated enhanced convergence speed and reliability for 3D motion planning, avoiding local optima through probabilistic state superposition \cite{du2024research}. In supply chain and route optimization contexts, variational quantum algorithms incorporating carefully crafted infeasible-solution constraints have been proposed to reduce qubit requirements, making near-term hardware usage more feasible \cite{li2024quantum}. Quantum generative models, such as QGANs and Quantum Boltzmann Machines, also offer pathways for realistic scenario synthesis, thus addressing robust planning in diverse weather conditions \cite{ramasamy2024enhancing}. Hybrid classical-quantum annealing solutions have been tested for generalized traveling-salesman-type routing, demonstrating potential speedups over purely classical solvers under certain problem setups \cite{hua2024quantum}. Additionally, QAOA-based formulations of multi-UAV path planning, encoded as quadratic unconstrained binary optimization (QUBO) problems, have shown promise for efficiently handling large search spaces, though challenges remain in scaling to industrial use cases \cite{davies2024quantum}.

\section{Problem Formulation\label{sec22}}
We consider a drone navigating in a two-dimensional continuous environment $\mathcal{E}\subseteq \mathbb{R}^2$ populated by a set of obstacles $\{ \Omega_i \}_{i=1}^m$, where each obstacle $\Omega_i \subset \mathcal{E}$ is represented by a closed region (e.g., a polygon). The drone is tasked with traveling from a start location $s \in \mathcal{E}$ to a target location $t \in \mathcal{E}$, subject to collision-avoidance constraints. Formally, any trajectory $\mathbf{x}(t)\!:\![0,T]\!\to\!\mathcal{E}$ must satisfy 
$\mathbf{x}(0) = s, \quad \mathbf{x}(T) = t, \quad
\text{and} \quad  \mathbf{x}(\tau) \notin \Omega_i, 
\text{where} \quad \forall \tau \in [0,T], \quad \text{and} \quad \forall i \in \{1, \dots, m\}.$

Additional kinematic constraints may be imposed, such as bounded velocities or turn rates, ensuring the path remains physically feasible for the drone.

The objective is to find a feasible path $\mathbf{x}(\cdot)$ that minimizes a cost functional reflecting travel distance, energy consumption, or flight time. A common formulation is expressed as $\min_{\mathbf{x}(\cdot)} \int_{0}^{T} c\Bigl(\mathbf{x}(\tau), \dot{\mathbf{x}}(\tau)\Bigr)\, d\tau$ 
$\text{subject to:} \quad 
\mathbf{x}(0) = s, \quad 
 \mathbf{x}(T) = t, \quad \text{and} \quad 
\mathbf{x}(\tau) \notin \bigcup_{i=1}^m \Omega_i, \quad \forall \tau \in [0,T],$
where $c(\cdot,\cdot)$ is a cost density that may incorporate travel distance, flight time, or risk factors (e.g., proximity to obstacles). Such path-planning problems are typically solved using either deterministic partial differential equation (PDE) approaches (e.g., fast marching methods) or Monte Carlo sampling-based techniques (e.g., Rapidly-exploring Random Trees). However, as the environment and obstacle set grow complex, or if multiple drones must be coordinated simultaneously, the computational complexity can become prohibitive. \textit{These challenges have driven research into alternative frameworks, including quantum and hybrid quantum-classical approaches, to accelerate the path-planning process while ensuring safety and collision avoidance. }

\section{Methodology \label{sec3}}
\begin{figure}[htpb]
    \centering
    \includegraphics[width=1\linewidth]{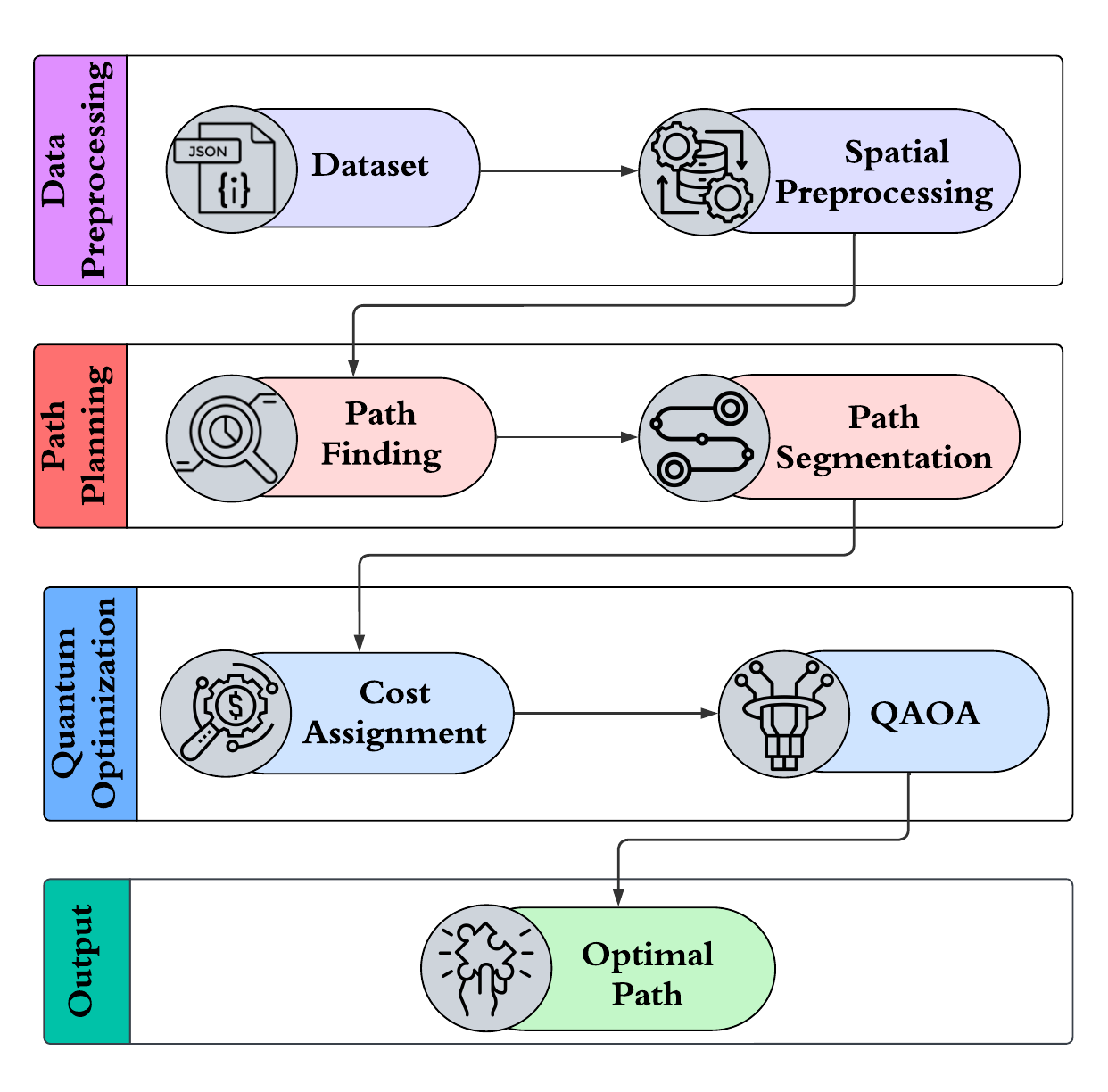}
    \vspace{-0.5cm}
\caption{  Overview of the proposed quantum-assisted path planning framework. The methodology consists of different main stages: \textbf{Data Preprocessing}, where the dataset undergoes spatial preprocessing; \textbf{Path Planning}, involving path finding and segmentation; \textbf{Quantum Optimization}, where cost assignment is followed by the application of QAOA to optimize the path; and \textbf{Output}, which yields the optimal path.}
    \label{fig:methodology}
\end{figure}

The proposed UAV path planning framework integrates classical spatial preprocessing with quantum optimization techniques to ensure efficient and collision-free navigation. As shown in Fig.~\ref{fig:methodology}, the methodology consists of three key phases: data processing, with spatial preprocessing as the main component; path planning, which includes pathfinding and path segmentation; and quantum-assisted optimization, which encompasses cost assignment and QAOA. The detailed steps are summarized in Algorithm \ref{alg:drone_path}.
\begin{algorithm}[htpb]
\caption{QUAV}
\label{alg:drone_path}
\footnotesize
\KwIn{Start point $\mathbf{P}_{\text{start}}$, end point $\mathbf{P}_{\text{end}}$, obstacle data.}
\KwOut{Optimized UAV path avoiding obstacles.}

\tcp{Spatial Preprocessing}
Convert GPS coordinates of $\mathbf{P}_{\text{start}}$, $\mathbf{P}_{\text{end}}$, and obstacles to UTM coordinates\;
Represent obstacles as polygons in UTM coordinates\;
Apply a safety buffer around each obstacle using scaling factors $(s_x, s_y)$\;

\tcp{Path Finding}
Discretize the space into a structured grid with resolution $r$\;
Construct a graph $G = (V, E)$ where $V$ are waypoints and $E$ are valid connections\;
Enumerate all candidate paths from $\mathbf{P}_{\text{start}}$ to $\mathbf{P}_{\text{end}}$\;
Compute Euclidean length and directional smoothness for each candidate path\;

\tcp{Path Segmentation}
Compute total Euclidean distance $D$\;
Determine segment size $\Delta s$ based on available quantum resources\;
Initialize $\mathbf{P}_{\text{current}} = \mathbf{P}_{\text{start}}$\;
\While{$\mathbf{P}_{\text{current}} \neq \mathbf{P}_{\text{end}}$}{
    Compute direction vector $\mathbf{v}_{\text{dir}}$\;
    Compute next waypoint $\mathbf{P}_{\text{next}}$\;
    Add edge $[\mathbf{P}_{\text{current}}, \mathbf{P}_{\text{next}}]$ to the path\;
    Update $\mathbf{P}_{\text{current}} = \mathbf{P}_{\text{next}}$\;
}

\tcp{Cost Assignment}
\For{each edge $e_i$ in the segmented path}{
    Compute Euclidean distance cost $C_{\text{dist}}(e_i)$\;
    Compute obstacle penalty $C_{\text{obs}}(e_i)$ based on buffer distance\;
    Assign final cost\footnotemark:
    \[
    C(e_i) = 
    \begin{cases}
        -10^3, & e_i \text{ is the start segment}, \\
        10^6, & e_i \text{ intersects an obstacle}, \\
        C_{\text{dist}}(e_i) + C_{\text{obs}}(e_i), & \text{otherwise}.
    \end{cases}
    \]
}

\tcp{Quantum-Assisted Optimization}
Map each path segment to a qubit\;
Initialize qubits in the equal superposition state\;
Apply cost Hamiltonian $H_C$\;
Apply mixer Hamiltonian $H_B$\;
Perform iterative optimization over $k$ layers\;
Optimize parameters $\gamma$ and $\beta$ using Adam optimizer\;

\KwResult{Optimized path as a set of edges.}

\end{algorithm}
\footnotetext{The negative cost assigned to the start segment serves as a bias to ensure that the path always begins at the designated start location. Its magnitude is negligible compared to obstacle penalties, so it does not distort the optimization objective.}
\subsection{Spatial Preprocessing}

The input data consists of the start point, end point, and obstacle data represented as GPS coordinates. Accurate spatial computations are critical for path planning; therefore, all coordinates are converted into Universal Transverse Mercator (UTM) coordinates using the transformation 
$    (\text{UTM}_{E}, \text{UTM}_{N}) = f(\phi, \lambda),$ where $\phi$ and $\lambda$ represent latitude and longitude, respectively, and $(\text{UTM}_{E}, \text{UTM}_{N})$ denote the projected easting and northing coordinates in meters. The projection function $f$ is implemented using the \texttt{pyproj} library. 

Obstacles are represented as polygons in the UTM coordinate system. To ensure safety, a buffer is applied around each obstacle, increasing its effective size
$    O_b = \text{scale}(O, s_x, s_y),
\label{eq:obstacle_buffer}, $
where $O$ is the original obstacle polygon, and $s_x, s_y > 1$ are scaling factors applied along the $x$- and $y$-axes. This guarantees a predefined safety margin for the drone.
\subsection{Path Finding}

Path finding aims to generate an initial set of candidate trajectories from the start point to the destination without considering obstacle avoidance at this stage. The process begins by discretizing the search space into a structured grid with resolution $r$, spanning the entire navigable area between the start and end points. Each discrete point in the grid, defined as $(x_i, y_j)$, serves as a potential waypoint and is computed as 
$G = { (x_i, y_j) | x_i = x_0 + i r, y_j = y_0 + j r, i, j \in \mathbb{Z} }.$

Once the grid is established, a graph representation is constructed where each waypoint corresponds to a node, and valid connections between adjacent waypoints define the edges. The connectivity of the graph can be determined using a 4-neighbor or 8-neighbor scheme, where in the former, each node connects to its immediate neighbors in the North, South, East, and West directions, and in the latter, additional diagonal connections are allowed. The set of nodes $V$ and edges $E$ forms the search graph $G = (V, E)$.

With the grid and graph structure in place, a systematic search is performed to generate candidate paths from the start node $s$ to the destination node $d$. Each trajectory is represented as a sequence of connected waypoints 
$P_k = { p_1, p_2, \dots, p_n },$
where $p_1 = s$ and $p_n = d$. The total set of candidate paths is represented as
$P = { P_1, P_2, \dots, P_m }.$

Each path is stored as a sequence of edges $E_k$ connecting consecutive waypoints. At this stage, no filtering is applied to remove paths that may intersect with obstacles. The goal is to enumerate all potential paths that could connect the start and end points within the grid space. These paths may vary significantly in length and structure, with some being direct while others follow more circuitous routes depending on the underlying connectivity of the grid.

Each generated path is then evaluated based on intrinsic properties, focusing solely on path geometry. The total Euclidean length of a path $P_k$ is computed as 
$L(P_k) = \sum_{i=1}^{n-1} | p_{i+1} - p_i |.$
Additionally, to assess directional smoothness, the cumulative angular deviation along the path is calculated to quantify abrupt changes in trajectory. Paths that exhibit minimal deviations and maintain a steady directional trend are considered structurally more optimal. 
This path-finding stage ensures that a complete set of structured candidate trajectories is generated before considering obstacle avoidance and optimization in subsequent steps.
\subsection{Path Segmentation}

To enable efficient trajectory optimization, the UAV path is segmented into discrete waypoints. The Euclidean distance between the start and end points in UTM coordinates is computed as 
$    D = \| \mathbf{P}_{\text{end}} - \mathbf{P}_{\text{start}} \|,$
where $\mathbf{P}_{\text{start}}$ and $\mathbf{P}_{\text{end}}$ are the start and end coordinates, respectively. The path is divided into $N$ adaptive segments with a step size defined as
   $ \Delta s = \frac{D}{N},$
where $N$ is selected based on the available quantum resources.

At each step, the next waypoint $\mathbf{P}_{\text{next}}$ is computed using the normalized direction vector:
$    \mathbf{v}_{\text{dir}} = \frac{\mathbf{P}_{\text{end}} - \mathbf{P}_{\text{current}}}{\| \mathbf{P}_{\text{end}} - \mathbf{P}_{\text{current}} \|}, \quad
    \mathbf{P}_{\text{next}} = \mathbf{P}_{\text{current}} + \mathbf{v}_{\text{dir}} \cdot \Delta s.$
\subsection{Quantum-Assisted Optimization}

The QAOA is employed to optimize segmented UAV paths while ensuring obstacle avoidance. By mapping each path segment to a qubit, QAOA enables the quantum circuit to explore multiple path configurations simultaneously, offering the potential to reduce computational complexity compared to classical approaches. The optimization process is structured into three key stages: initialization, cost Hamiltonian application, and mixer Hamiltonian application, followed by iterative quantum-classical optimization.

The initialization stage prepares the qubits in an equal superposition state, ensuring that a broad set of possible paths is initially considered. Each qubit corresponding to a path segment is set in the state $|+\rangle$ through the application of Hadamard gates, such that the initial state is given by 
$    | \psi_0 \rangle = \bigotimes_{i=1}^{N} H |0\rangle.$
The cost Hamiltonian is designed to encode the optimization problem, incorporating penalties for inefficient paths and obstacle proximity: 
$H_C = \sum_{i=1}^{N} C(e_i) Z_i,$
where $C(e_i)$ represents the cost associated with path segment $e_i$, and $Z_i$ is the Pauli-Z operator acting on qubit $i$. The cost function penalizes paths that intersect obstacles, assigns an additional cost to paths passing within a predefined buffer distance from obstacles, and incorporates a term for path efficiency by considering segment length.
Additionally, a small negative bias is applied to the start segment to anchor the optimization at the correct starting point.
The corresponding quantum gate implementation applies a sequence of controlled-NOT (CNOT) gates, followed by a $R_z(2\gamma)$ rotation on the target qubit, and then another CNOT gate between the previous qubit and the current qubit. This ensures that dependencies between connected path segments are accounted for, preserving path continuity. The unitary transformation corresponding to the cost Hamiltonian is given by 
   $ U_C(\gamma) = e^{-i \gamma H_C}.$
   
The mixer Hamiltonian promotes exploration of alternative paths by applying Pauli-X rotations, allowing the system to transition between different path configurations. It is defined as 
  $  H_B = \sum_{i=1}^{N} X_i,$
where $X_i$ is the Pauli-X operator acting on qubit $i$. During the quantum evolution, this is implemented by applying an $R_x(2\beta)$ gate to each qubit, allowing transitions between different path states. The corresponding unitary transformation is given by 
   $ U_B(\beta) = e^{-i \beta H_B}.$

The circuit alternates between the application of the cost and mixer Hamiltonians for $k$ layers, progressively refining the solution. The resulting quantum state after $k$ iterations is given by 
$    | \psi(\gamma, \beta) \rangle = \prod_{p=1}^{k} U_B(\beta_p) U_C(\gamma_p) | \psi_0 \rangle.$

After execution, the quantum state is measured, and the resulting bitstrings represent candidate paths. These outcomes are fed into a classical optimizer, which iteratively updates the parameters $\gamma$ and $\beta$ to minimize the expectation value of the cost Hamiltonian:
$    \langle H_C \rangle = \langle \psi(\gamma, \beta) | H_C | \psi(\gamma, \beta) \rangle.$

The optimization framework ensures collision-free paths by incorporating an obstacle buffer region into the cost function. If a segment lies within a safety margin $d_s$ of an obstacle, an additional penalty term is introduced: 
   $ C(e_i) = C(e_i) + \lambda \cdot \exp(-d_i/d_s),$
where $d_i$ represents the distance from the segment to the nearest obstacle, and $\lambda$ is a scaling factor. This exponentially increasing penalty discourages paths that pass too close to obstacles, ensuring a safe trajectory. The adaptability of this framework allows the cost function to be fine-tuned to prioritize factors such as shortest distance, energy efficiency, or safety margins, depending on specific UAV mission requirements. 

\begin{figure}[htpb]
    \centering
    \includegraphics[width=1\linewidth]{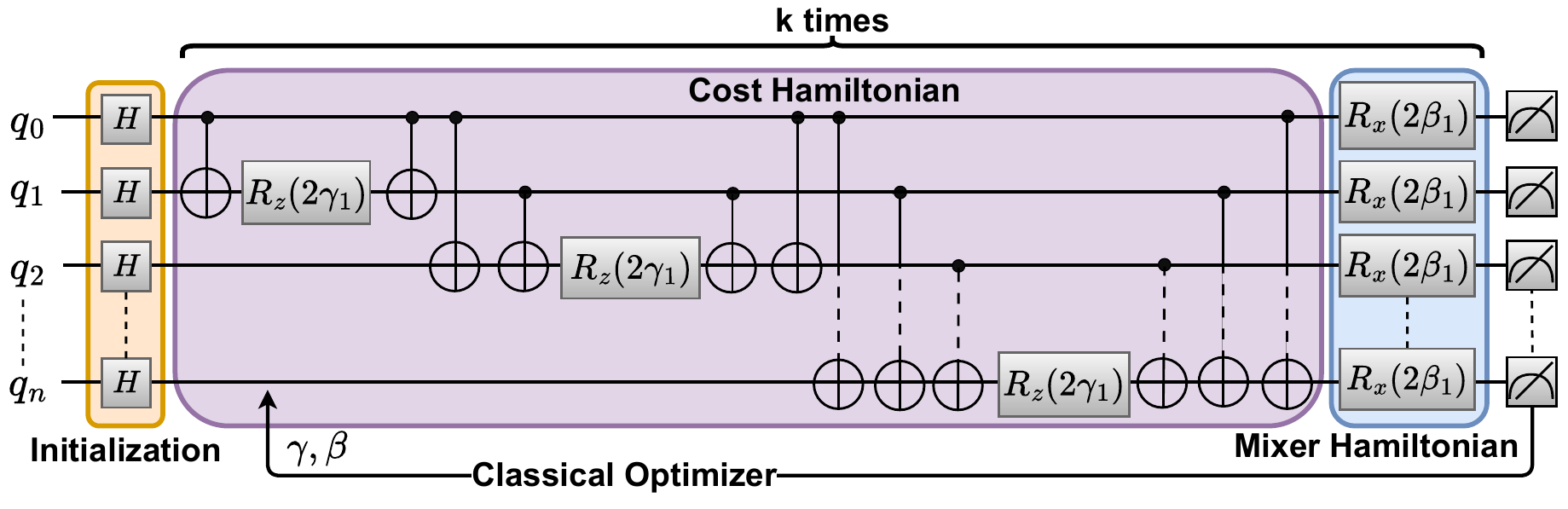}
    \vspace{-0.2cm}
\caption{Quantum circuit representation of the QAOA. The circuit consists of three main stages: (1) \textbf{Initialization}, where Hadamard gates prepare the initial uniform superposition state; (2) \textbf{Alternating Cost and Mixer Hamiltonians}, applied iteratively for \( k \) layers. Each iteration consists of a \textbf{Cost Hamiltonian}, which encodes the problem constraints through controlled phase rotations \( R_z(2\gamma) \), followed by a \textbf{Mixer Hamiltonian}, which applies \( R_x(2\beta) \) rotations to explore the solution space; and (3) \textbf{Measurement}, where the final quantum state is measured to extract the optimal solution. The parameters \( \gamma \) and \( \beta \) are optimized using a classical optimizer in a feedback loop to iteratively improve the solution.}
    \label{fig:qaoa}
\end{figure}

Fig.~\ref{fig:qaoa} illustrates the QAOA circuit, highlighting the initialization, cost, and mixer Hamiltonians, and the iterative quantum-classical optimization process.

\subsection{Time Complexity \label{35}}
To determine the computational complexity of the QUAV algorithm, we analyze each phase separately. The preprocessing step, which involves converting GPS coordinates to UTM and representing obstacles as polygons, scales with the number of obstacles $M$ and their average number of vertices $\bar{v}$, resulting in a complexity of $\mathcal{O}(M\bar{v})$.

Next, the pathfinding phase constructs a graph $G = (V, E)$, where $V$ represents waypoints and $E$ the valid edges. In structured grids, each waypoint is connected to a fixed number of neighbors, leading to a total number of edges $|E| = \mathcal{O}(|V|)$. After graph construction, path segmentation is performed by dividing the total path into discrete edges, which requires at most $\mathcal{O}(|E|)$ operations.

For optimization, each path segment (edge) is mapped to a qubit, so the number of qubits is $n = |E|$. In the quantum-assisted optimization phase, QAOA applies a cost Hamiltonian and a mixer Hamiltonian iteratively over $k$ layers. Under the assumption of sparse 2-local interactions without additional ancilla qubits, each layer consists of $\mathcal{O}(|E|)$ quantum gates. Parameter training requires $S$ classical iterations, and each iteration evaluates the cost function using a fixed number of measurement shots $R$. The complexity of the quantum optimization phase is therefore
$T_{\text{QAOA}} = \mathcal{O}(S \cdot R \cdot k |E|).$

Summing all components, the overall complexity of QUAV is expressed as
$T_{\text{total}} = \mathcal{O}(M\bar{v} + |E| + S \cdot R \cdot k |E|).$

Under fixed shot budgets $R$, and constant circuit depth $k$, the asymptotic complexity is governed by
$T_{\text{total}} = \mathcal{O}(S \cdot |E|).$

Thus, QUAV exhibits linear scaling in circuit depth with respect to the number of QAOA layers and the number of qubits (equal to the number of edges in our formulation), while the number of optimization steps $S$ remains the main factor governing practical runtime. This reflects circuit resource scaling; in practice, the classical optimization of QAOA may introduce additional overhead that can scale more severely with system size.
\section{Results \label{sec4}}
\subsection{Experimental Settings}

The key experimental settings used for the drone path planning optimization are summarized in Table~\ref{tab:exp_settings}. Various configurations are tested, and these final settings are determined as the best-performing ones.

\begin{table}[htpb]
    \centering
    \caption{Experimental settings.}
    \scriptsize
    \label{tab:exp_settings}
    \renewcommand{\arraystretch}{1.2} % Adjust row height for readability
    \begin{adjustbox}{max width=0.9\linewidth}
    \begin{tabularx}{\linewidth}{|X|X|}
        \hline
        \rowcolor{purple!20} \textbf{Parameter} & \textbf{Value} \\
        \hline
        \textbf{Quantum Device} & \texttt{default.qubit} (PennyLane) and \texttt{ibm\_kyiv} (IBM QPU) \\
        \hline
        \textbf{Qubits} & 20 (1 per edge) \\
        \hline
        \textbf{Depth [Layers $k$]} & 5 \\
        \hline
        \textbf{Optimizer} & Adam \\
        \hline
        \textbf{Learning Rate} & 0.1 \\
        \hline
        \textbf{Optimization Steps} & 60 (20 for QPU) \\
        \hline
        \textbf{Start and End Points} & 6 Scenarios \\
        \hline
        \textbf{Coordinate Projection} & GPS to UTM (Zone 49) \\
        \hline
        \textbf{Step Size} & Total Distance / 20 \\
        \hline
        \textbf{Obstacle Buffer} & 5 meters \\
        \hline
    \end{tabularx}
    \end{adjustbox}
\end{table}

We use the \texttt{default.qubit} simulator from PennyLane for initial testing and prototyping \cite{bergholm2018pennylane}. For real quantum executions, the IBM QPU \texttt{ibm\_kyiv} is used through Qiskit \cite{javadi2024quantum}, allowing validation on actual quantum hardware. 

For the QAOA setup, 20 qubits are assigned, each representing an edge in the path. The Adam optimizer is employed with a learning rate of 0.1 to ensure stable convergence. Optimization is performed in 60 steps, with 20 allocated specifically for QPU execution to balance computational load and solution quality.

Experiments cover six different start and end point scenarios, each with varying complexities. GPS coordinates are projected into UTM (Zone 49) using PyProj for accurate distance calculations \cite{Pyproj}. The step size is dynamically set as the total distance divided by 20 to match the qubit count. %A 5-meter buffer is applied around obstacles using Shapely to maintain a safe drone path.

To assess the effectiveness of the proposed quantum approach, we benchmark performance using two primary metrics: path distance and loss values. The path distance metric evaluates the efficiency of the computed trajectory by measuring the total distance traveled by the drone, ensuring it remains as short as possible while avoiding obstacles. The loss function is defined based on the path feasibility and smoothness, penalizing trajectories that pass too close to obstacles or exhibit excessive deviation from optimality.

Since this work represents the first exploration of quantum-based drone pathfinding and our primary objective is to validate the fundamental viability of a QAOA-based planner rather than to outperform mature classical methods, we benchmark QUAV against two widely used baseline algorithms, A* (A-star) and Rapidly-exploring Random Tree (RRT) \cite{hart1968formal,xu2024recent,lavalle1998rapidly}. Although more advanced classical planners and hybrid variants exist, aligning our evaluation with these foundational techniques provides a clear reference for assessing path quality and obstacle-avoidance behavior under identical geospatial and efficiency criteria. The hyperparameters for these baseline methods are chosen based on standard configurations in the literature:
\begin{itemize}
    \item \textbf{A*}: Euclidean distance heuristic, grid resolution of $0.5$ meters, and a smoothing factor of $0.2$.
    \item \textbf{RRT}: Step size of $1.0$ meters, maximum iterations of $1000$, and a goal bias of $0.05$.
\end{itemize}
\textit{These classical methods serve as a reference to validate the feasibility and practical applicability of our quantum-enhanced optimization without making direct claims of superiority over classical techniques.}

\subsection{Loss Analysis}

\begin{figure}[htpb]
    \centering
    \includegraphics[width=1\linewidth]{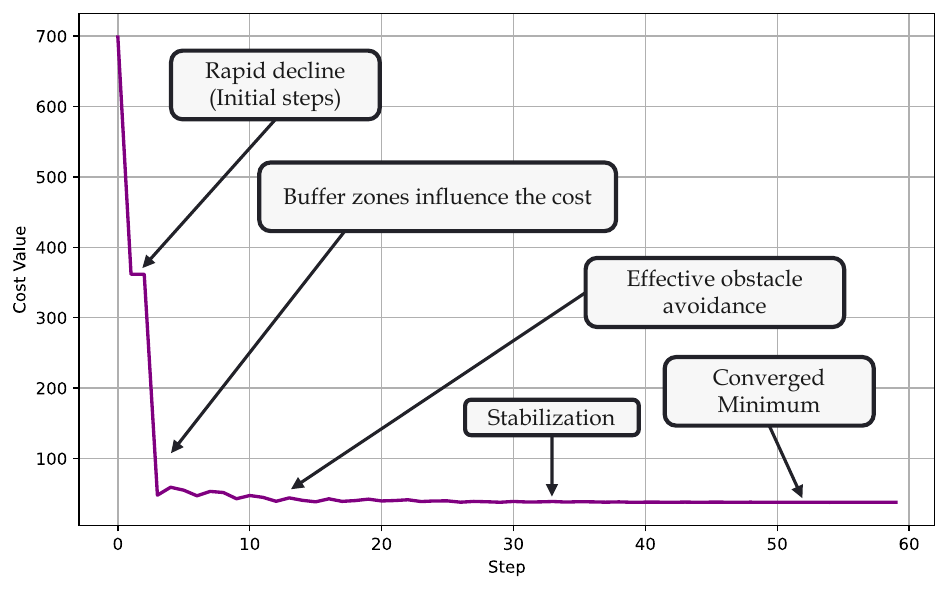}
    \vspace{-0.4cm}
\caption{Cost function convergence during the optimization process using default simulator [PennyLane]. The curve demonstrates an initial \textbf{rapid decline} in cost, followed by a phase where \textbf{buffer zones influence the cost}, leading to \textbf{stabilization}. Further optimization ensures \textbf{effective obstacle avoidance}, eventually reaching a \textbf{converged minimum}. This behavior highlights the efficiency of the quantum-assisted optimization in refining the path while ensuring obstacle-free navigation.}
    \label{fig:loss}
\end{figure}

The first evaluation of our results focuses on the loss convergence curve, which illustrates how the cost evolves over the optimization steps. Fig.~\ref{fig:loss} presents the loss function's behavior throughout the training process.

\begin{figure}[htpb]
    \centering    \includegraphics[width=1\linewidth]{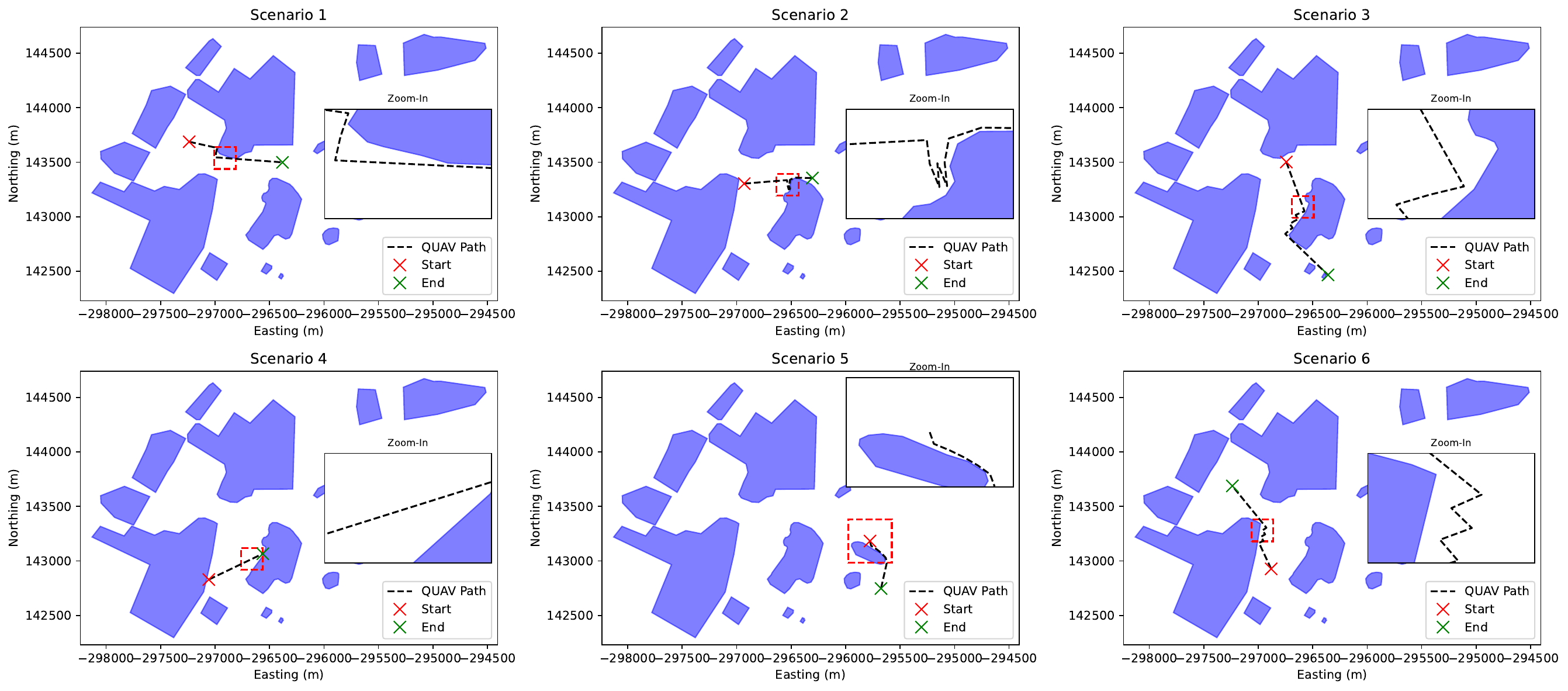}
    \vspace{-0.2cm}
\caption{Obstacle avoidance performance across six scenarios using QUAV. The blue-marked regions represent obstacles, defining restricted areas that the path must avoid. The dashed black line illustrates the computed QUAV path, navigating through the environment while minimizing deviations. Red and green markers indicate the start and end points, respectively. Each scenario includes a zoomed-in view highlighting key avoidance maneuvers.}
\label{quav}
\end{figure}
At the beginning of the optimization, a rapid decline in the cost is observed during the first 5 to 10 steps. This sharp drop is expected as the optimizer quickly adjusts the parameters $\gamma$ and $\beta$ to eliminate inefficient paths, particularly those that either collide with obstacles or pass too close to them. In this phase, the optimizer aggressively penalizes paths that violate the buffer constraints.

Following this initial drop, the curve enters a stabilization phase between steps 10 and 40. During this phase, the optimizer fine-tunes the parameters to balance between minimizing the path length and maintaining a safe distance from obstacles. This phase is crucial for ensuring that the generated path remains both feasible and compliant with the 5-meter obstacle buffer.

Finally, around step 50, the loss curve converges to a minimum, indicating that the optimizer has found an optimal or near-optimal solution. The cost function stabilizes, suggesting that further iterations yield diminishing returns.
\subsection{Obstacle Avoidance Analysis}

Now, we analyze the performance of our approach in terms of its ability to effectively avoid obstacles across six different scenarios. Fig.~\ref{quav} visualizes the UAV's trajectory in each scenario, demonstrating how the algorithm ensures collision-free navigation while progressing toward the target.

In each scenario, the UAV is tasked with navigating from the start point to the end point while avoiding the blue-marked obstacles. The QUAV path is represented by the dashed black line. Across all six scenarios, the algorithm consistently demonstrates its ability to avoid obstacles while still progressing toward the destination. Even in densely populated areas, as highlighted in the zoom-in views, the UAV successfully navigates through narrow corridors without intersecting any obstacles, ensuring collision-free movement.

In several cases, such as Scenarios 3 and 6, a zigzag pattern is observed instead of a straight path. This behavior arises due to the probabilistic nature of QAOA, where the algorithm favors lower-cost paths that balance safety and efficiency but may not always select the most direct route. The cost function strongly penalizes proximity to obstacles, sometimes favoring longer but safer trajectories. Due to the probabilistic sampling in QAOA, the optimizer explores multiple valid paths with similar costs, occasionally selecting paths with more complex geometries. Despite these deviations, the paths remain optimal within the defined constraints, ensuring obstacle-free navigation.

Another contributing factor to the observed zigzag patterns is the inherent probabilistic sampling in quantum algorithms like QAOA. The solution space often contains multiple valid paths with nearly equivalent costs. As a result, the optimizer may select paths that, while feasible, exhibit additional complexity. However, in cases such as Scenario 4, the path remains relatively smooth, indicating that the algorithm identified a more direct and cost-effective route.

Despite variations in path complexity, a key takeaway is that in all six scenarios, \textit{the UAV successfully reaches its destination while maintaining a collision-free path}. The algorithm demonstrates adaptability across different obstacle layouts, effectively applying cost penalties to ensure safe navigation. Whether operating in tight spaces or open areas, the optimization consistently avoids obstacles while optimizing the flight trajectory.

\begin{figure}[htpb]
    \centering
    \includegraphics[width=1\linewidth]{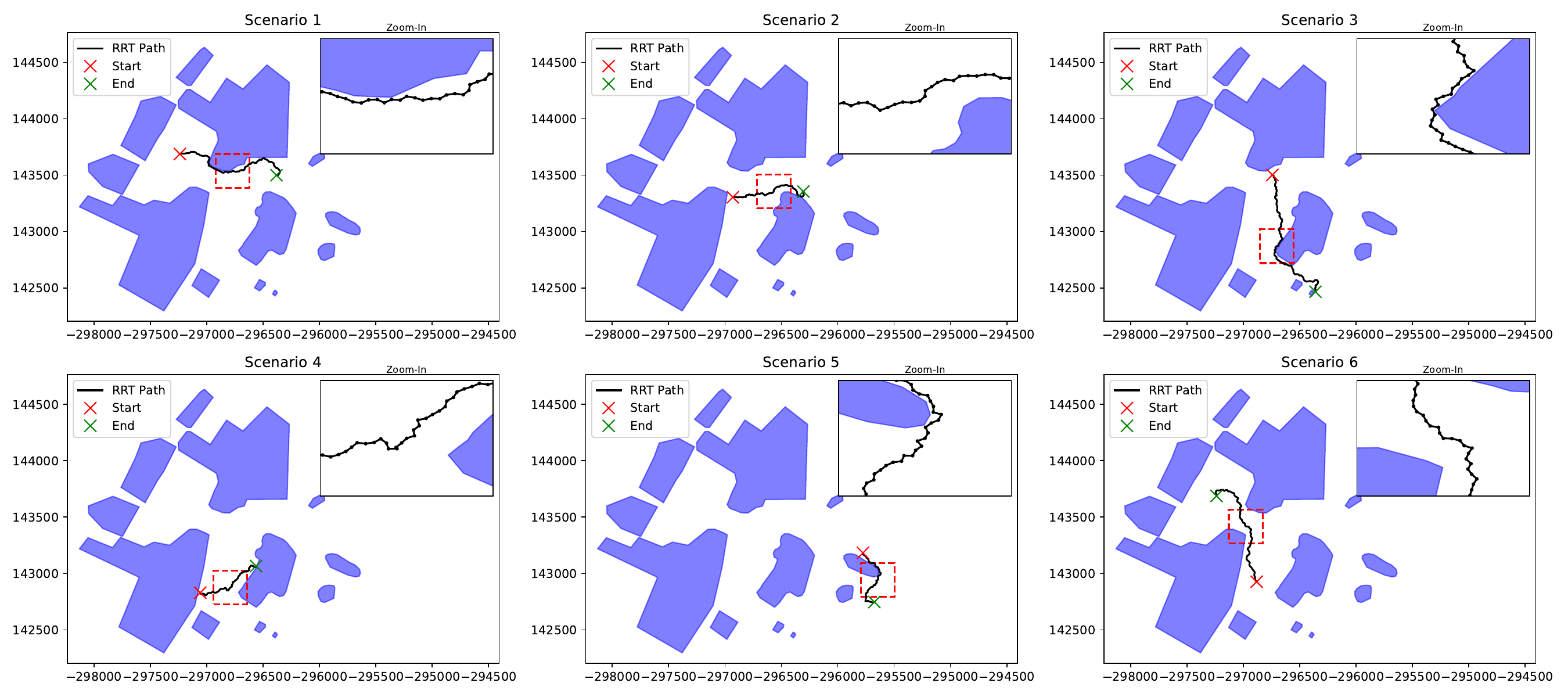}
    \vspace{-0.2cm}
\caption{Obstacle avoidance performance across six scenarios using RRT. The generated paths exhibit exploratory characteristics, often leading to non-optimal deviations due to the stochastic nature of tree expansion.}
    \label{fig:rrt_path}
\end{figure}
\begin{figure}[htpb]
    \centering
    \includegraphics[width=1\linewidth]{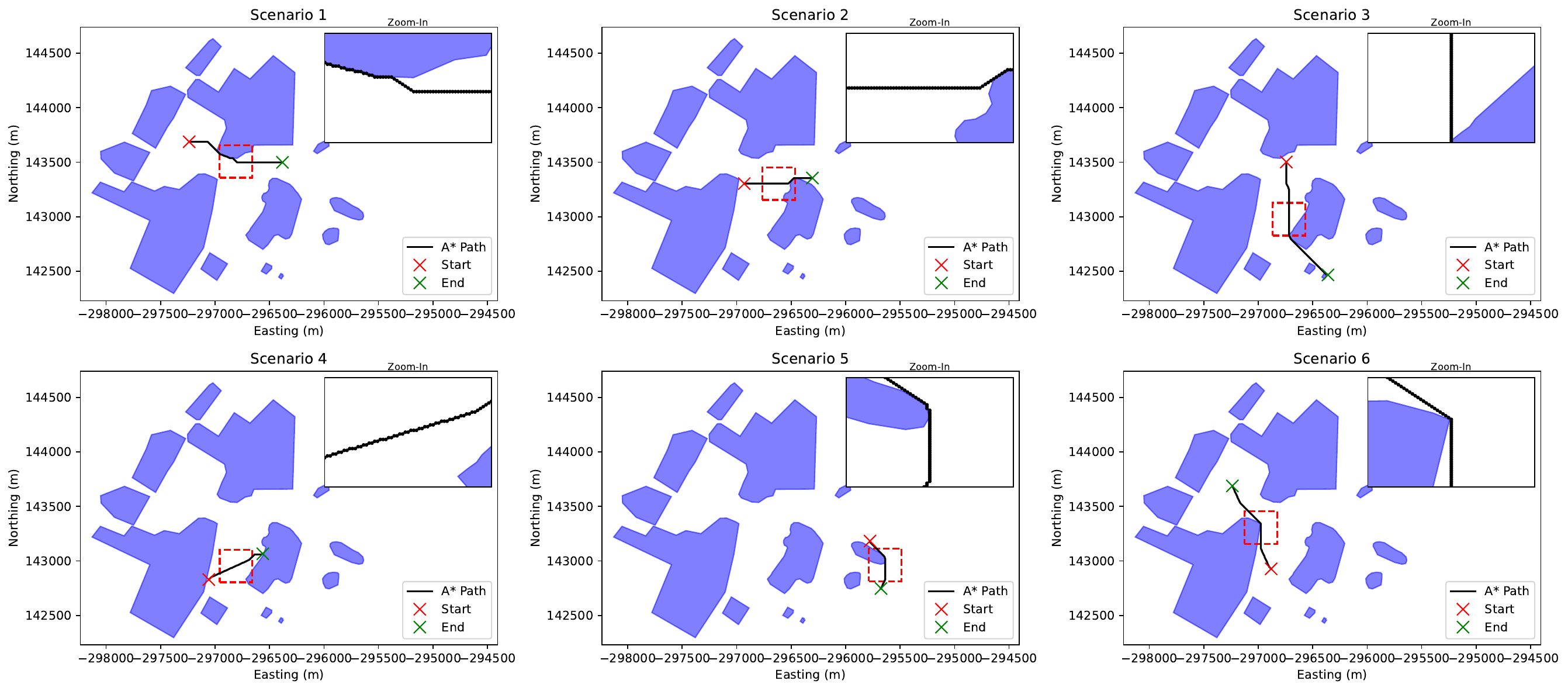}
    \vspace{-0.2cm}
\caption{Obstacle avoidance performance across six scenarios using A*. The computed paths prioritize optimality, leveraging heuristic-based search to determine the shortest, collision-free trajectory. }
    \label{fig:astar_path}
\end{figure}
Figs~\ref{fig:rrt_path} and~\ref{fig:astar_path} illustrate how RRT and A* navigate around obstacles. A* finds the shortest path while strictly avoiding obstacles, whereas RRT generates a more exploratory trajectory, sometimes requiring additional smoothing to refine the final path. 

\subsection{Distance Analysis}

To evaluate the effectiveness of QUAV in path optimization, we compare the path lengths obtained by A*, RRT, and QUAV across six different scenarios. Table~\ref{tab:distance} presents the path distances for each method.

\begin{table}[htpb]
    \centering
    \caption{Path distance comparison across six scenarios (S1-S6).}
    \scriptsize
    \label{tab:distance}
    \begin{adjustbox}{max width=1\linewidth}
    %\scriptsize
    \begin{tabular}{|c|c|c|c|c|c|c|}
        \hline
        \rowcolor{purple!20} \textbf{Algorithm} & \textbf{S1} & \textbf{S2} & \textbf{S3} & \textbf{S4} & \textbf{S5} & \textbf{S6} \\
        \hline
        \textbf{A*} & 855.00 & 625.00 & 380.00 & 495.00 & 380.00 & 655.00 \\
        \hline
        \textbf{RRT} & 1128.65 & 825.12 & 1561.99 & 703.52 & 662.88 & 1130.38 \\
        \hline
        \textbf{QUAV} & 876.86 & 627.97 & 1104.11 & 548.73 & 447.08 & 976.88 \\
        \hline
    \end{tabular}
    \end{adjustbox}
\end{table}

The A* algorithm consistently delivers the shortest path across all scenarios, which is expected given its nature as an optimal pathfinding algorithm. However, A* suffers from scalability issues, as we will discuss in the time complexity analysis.

QUAV, our quantum-assisted approach, consistently outperforms RRT in finding shorter paths while remaining computationally efficient. For example, in Scenario 5, QUAV finds a path of 447.08 meters compared to RRT's 662.88 meters, demonstrating a significant improvement. While QUAV does not always match A* in absolute path optimality, it generally offers a more scalable alternative while still maintaining shorter paths than RRT. An exception is Scenario 3, where QUAV produces a path of 1104.11 meters, much longer than A* (380.00 meters) but still an improvement over RRT (1561.99 meters). These deviations are influenced by the probabilistic nature of QAOA, which favors paths with lower costs but may introduce variations in trajectory length.

\subsection{Time Complexity Analysis}
Time complexity is crucial for scalable pathfinding in real-time applications such as robotics and autonomous navigation. Table~\ref{tab:time_complexity} compares the computational complexities of A*, RRT, and QUAV.

\begin{table}[htpb]
    \centering
    \caption{Time complexity of pathfinding algorithms.}
    \scriptsize
    \label{tab:time_complexity}
    \begin{adjustbox}{max width=2\linewidth}
    \begin{tabular}{|p{4cm}|p{4cm}|}
        \hline
        \rowcolor{purple!20} \hfil \textbf{Algorithm} & \hfil\textbf{Complexity} \\
        \hline
       \hfil \textbf{A*} &\hfil  $\mathcal{O}(b^d)$ \\
        \hline
        \hfil\textbf{RRT} & \hfil$\mathcal{O}(|E|\log|E|)$ \\
        \hline
        \hfil\textbf{QUAV} & \hfil $\mathcal{O}(S \cdot |E|)$ \\
        \hline
    \end{tabular}
    \end{adjustbox}
\end{table} 
A* offers optimality but can, in the worst case, require exponential time $\mathcal{O}(b^d)$ \cite{stuartl2021artificial,kuprevsak2024pathfinding}, 
where $b$ denotes the branching factor of the search tree and $d$ is the depth of the solution. On uniform grids where each node has constant degree  ($b=\Theta(1)$), the solution depth is $d=\Theta(\sqrt{|V|})$. Since $|E|=\Theta(|V|)$ in such grids, the worst-case number of expansions grows as $\exp(\Theta(\sqrt{|E|}))$. This limits its applicability in large or high-dimensional graphs.

RRT scales better at $\mathcal{O}(N \log N)$ through randomized exploration \cite{karaman2011sampling}, where $N$ is the number of sampled nodes. In our grid-based formulation, the number of samples is proportional to the number of valid edges, so this complexity can also be expressed as $\mathcal{O}(|E|\log |E|)$. However, the stochastic nature of RRT results in suboptimal path quality, often requiring additional post-processing techniques such as smoothing or re-optimization to generate feasible and efficient paths. Its improved computational scalability makes RRT a viable choice for real-time applications, albeit with a compromise in optimality.

QUAV achieves $\mathcal{O}(S \cdot |E|)$ complexity (Sec.~\ref{35}), with circuit depth scaling linearly with both the number of QAOA layers and the number of qubits. Under fixed shot and evaluation budgets, this provides scalable performance, although the number of optimization steps $S$ remains the main practical factor. While current implementations emphasize 
feasibility rather than outperforming classical methods, integrating techniques such as Grover's algorithm \cite{grover1996fast} may further improve efficiency. QUAV's quantum-based design, therefore, offers a promising direction for large-scale, dynamic environments as quantum hardware advances.

\subsection{Hardware Results}

\begin{figure}[htpb]
    \centering
    \includegraphics[width=1\linewidth]{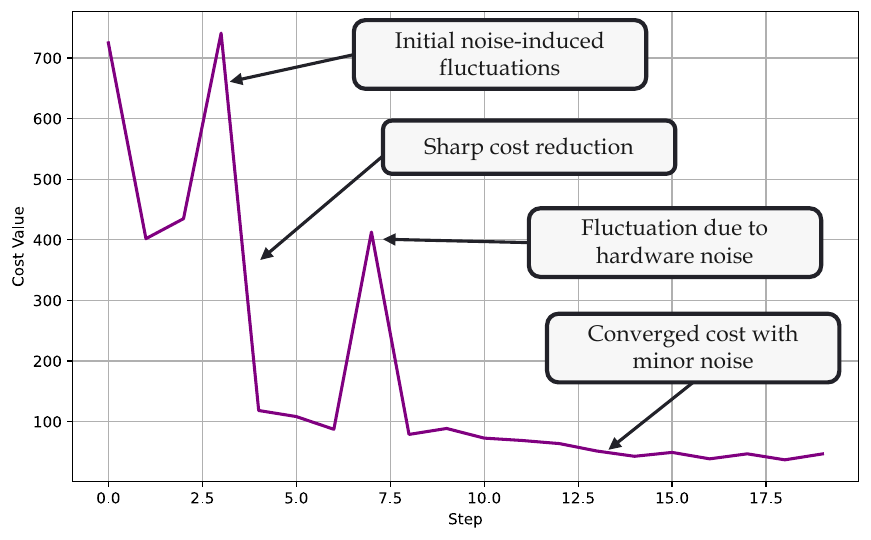}
    \vspace{-0.4cm}
\caption{ Cost function convergence during the optimization process using the IBM \texttt{ibm\_kyiv} QPU. The plot illustrates \textbf{initial noise-induced fluctuations} due to quantum hardware imperfections, followed by a \textbf{sharp cost reduction}. Subsequent \textbf{fluctuations} arise from \textbf{hardware noise}, before the cost function \textbf{converges with minor residual noise}. This behavior highlights the impact of quantum noise on the optimization process and the challenges of executing variational quantum algorithms on near-term quantum hardware.}
\label{fig:hardware_loss}
\end{figure}

The hardware results are obtained using IBM Quantum's QPU, specifically, the \texttt{ibm\_kyiv} backend. Fig.~\ref{fig:hardware_loss} presents the loss convergence behavior during the optimization process executed directly on the quantum hardware.

One of the first notable aspects of the hardware execution is the increased variability in cost values compared to the simulated runs. This variability arises due to hardware noise, decoherence, and readout errors, which are inherent in today's NISQ devices. In the early optimization steps, significant fluctuations appear, particularly a sharp spike around step 3, likely caused by hardware-induced errors affecting the optimization landscape. Despite these challenges, the optimizer effectively reduces the cost after approximately step 5, and the values stabilize towards the later steps, demonstrating that the algorithm can still find lower-cost paths even in the presence of quantum noise.

\begin{figure}[htpb]
    \centering
    %\includesvg{topology.svg}
    \includegraphics[width=0.85\linewidth]{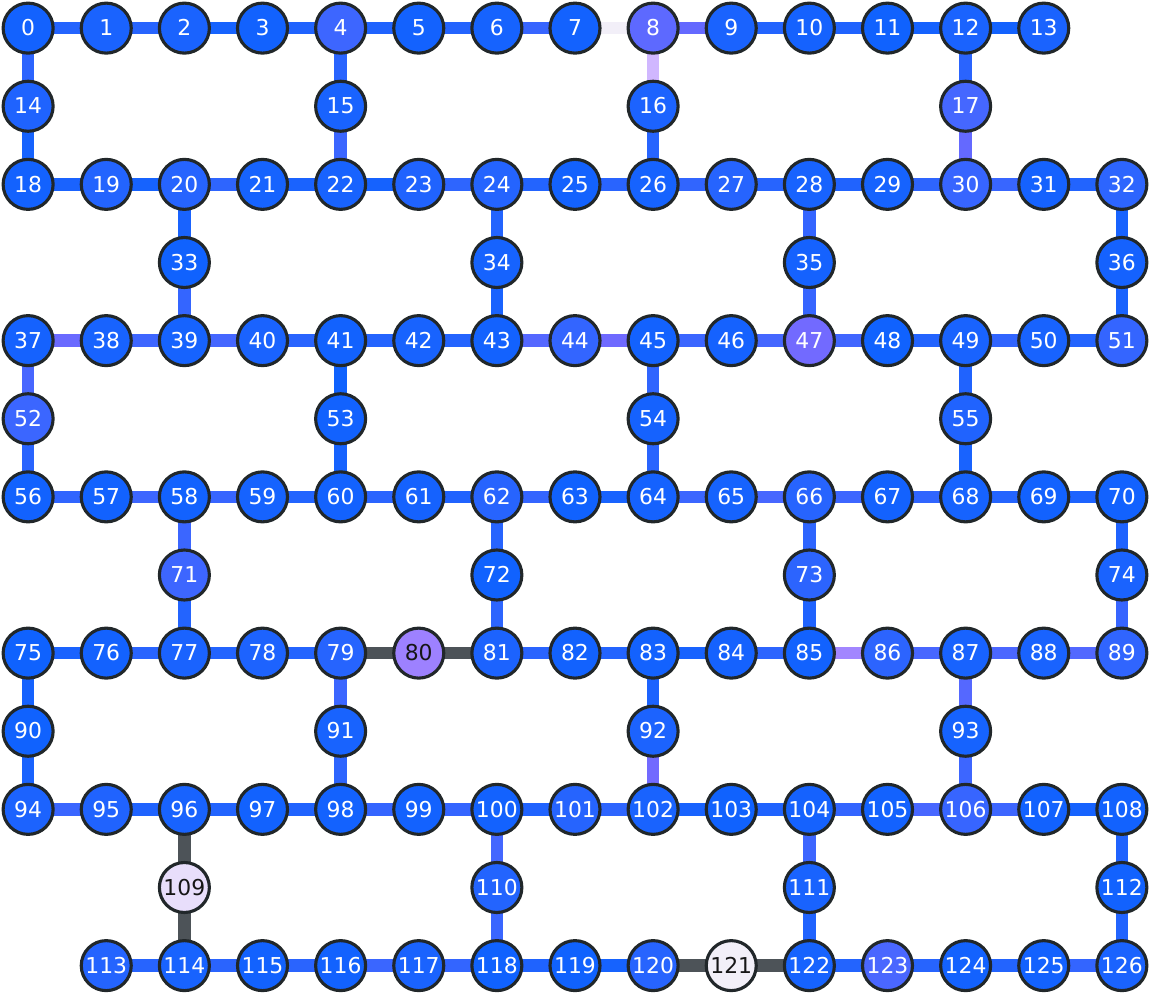}
    \vspace{0.2cm}
    \caption{Qubit connectivity map of the IBM \texttt{ibm\_kyiv} QPU. Each node represents a physical qubit, and edges show available couplings.}
    \label{fig:qpu_topology}
\end{figure}

The execution on real hardware is also influenced by the connectivity constraints of the QPU, as shown in Fig.~\ref{fig:qpu_topology}, which illustrates the topology of the \texttt{ibm\_kyiv} processor. Each node represents a qubit, and edges indicate available couplings for two-qubit operations. Mapping the logical qubits from the algorithm onto the physical qubits requires careful optimization, as limited connectivity can introduce additional SWAP operations, increasing error rates. Some qubits exhibit higher noise levels or lower fidelity, further impacting performance. Addressing these constraints through optimized transpilation and error mitigation strategies is crucial for enhancing the reliability of QAOA-based optimization on real quantum devices.

\section{Discussion \label{sec5}}
QUAV demonstrates a balance between computational efficiency and path quality, adapting well to constrained environments. Its convergence behavior, characterized by initial fluctuations followed by stabilization, reflects QAOA's probabilistic nature, enabling the flexible exploration of feasible paths rather than deterministic optimization.
In obstacle-rich settings, QUAV maintains safety while favoring efficient routes. While some zigzag patterns emerge, they reflect trade-offs between path length and constraint avoidance, offering robustness where classical methods may falter.
Scalability is evident, with increased qubit counts enhancing path quality through better solution sampling. However, real hardware introduces noise and connectivity issues, leading to cost fluctuations. These challenges underscore the importance of error mitigation and efficient qubit mapping.
Hardware-aware execution also reveals the need for optimized transpilation, as limited connectivity increases gate overhead. Despite this, QUAV retains convergence, suggesting resilience even under NISQ constraints.
While full quantum advantage remains theoretical, combining QAOA with classical preprocessing or Grover-based amplification offers a promising path forward. QUAV exemplifies quantum path planning's potential, distinct from classical paradigms, with growing viability as hardware matures.
\section{Conclusion \label{sec6}}
We introduced QUAV, a QAOA-based planner that balances path quality and scalability. Compared to A* and RRT, it offers a trade-off: faster runtime than A*, and higher-quality paths than RRT.
Experiments on quantum hardware reveal noise-induced cost variability but confirm convergence. Hardware-aware strategies are essential for reliable execution.
Despite current limitations, QUAV shows promise for real-world navigation. Future improvements in quantum hardware, error mitigation, and hybrid methods could further enhance its impact.
Although our goal at this stage is not to outperform classical methods, this work establishes QUAV as a practical first step toward scalable quantum-assisted path planning, with the potential to complement and eventually surpass classical approaches as quantum hardware matures.

 \section*{Acknowledgment}
 This work was supported in part by the NYUAD Center for Quantum and Topological Systems (CQTS), funded by Tamkeen under the NYUAD Research Institute grant CG008.
%{
   % \footnotesize
  
\bibliographystyle{IEEEtran}

\bibliography{refs}
%}
%\end{spacing}

\end{document}